# Superlubricity-Based Electrostatic Microgenerators

Xuanyu Huang[1], Li Lin[1], Quanshui Zheng[1,2*]

[1] Department of Engineering Mechanics and Center for Nano and Micro Mechanics, Tsinghua University, Beijing 100084, China
[2] State Key Laboratory of Tribology and Applied Mechanics Lab, Tsinghua University, Beijing 100084, China
*E-mail: zhengqs@tsinghua.edu.cn

**Although electrostatic microgenerators (ESMGs) have promised for nearly two decades extensive applications in wireless, self-powered microscale devices and sensors for security, personal health systems, communication, infrastructure and environmental monitoring [1, 2, 3, 4], commercialized ESMGs are still scarce, mainly due to very low current densities and short product life for most applications [1, 5, 6, 7, 8]. Here we demonstrate that a combination of structural superlubricity, a state of nearly zero friction and wear between two contacted solid surfaces [9, 10, 11, 12, 13, 14], and nanotechnology can endow ESMGs with superlong life and high performances such as transferred charge, current, and power (at least three orders of magnitude higher than those of conventional ESMGs). Among a few significant advantages of superlubricity-based ESMGs, two are particularly noticeable. First, they can be driven by much lower external loads than conventional ESMGs; second, they can be massively fabricated by using micro-machining processes. The quantitative relationships, experimental proof of concept, and optimization results reported in this Letter can guide future design and facilitate the commercialization of ESMGs.**

With the rapid development of nanotechnology and micro-machining processes, microscale sensors and devices are emerging in large numbers of applications in security, personal health systems, communication, infrastructure and environmental monitoring [1, 2, 3, 4]. These sensors and devices are mostly powered by batteries and external chargers, which seriously limits their further applications [15]. Especially for micro-/nano-electromechanical systems working in special conditions, such as micro-robots working in human blood vessels [16], it is very inconvenient to supply power from outside. It would be ideal if microgenerators could be individually installed in these systems and harvest energy from the environment [1, 2, 7]. Therefore, microgenerators promise extensive applications.

There are mainly three types of microgenerators, based on electromagnetic, electrostatic, and piezoelectric principles, respectively. In general, electromagnetic conversion is more suited for large systems because its efficiency scales somehow between $L^3$ and $L^4$, piezoelectric conversion is somehow better on the mesoscale, while electrostatic conversion becomes superior once devices are further miniatured because its efficiency scales somehow between $L$ and $L^2$ [17]. Thus, it is obvious why commercialized ESMGs are scarce.

In fact, quite a few types of electrostatic generators, such as triboelectric nanogenerator (TENG) [18] and electret-based microgenerator (EBMG) [19], have been proposed and intensively investigated during the last decade. The concept of TENGs is particularly elegant because TENGs are easy to charge just through contact. However, reported TENGs seem to be all in the macroscale, and no EBMGs have been commercialized. The major obstacles to the invention of commercialized microscale TENGs and EBMGs are their low output performances and short product life for most applications, caused mainly by friction and wear in principle. In order to reduce the influence of friction, various efforts have been made, such as using rolling friction [20, 21] or liquid friction [22] instead of direct contact sliding. However, these efforts may result in not only a more complicated structure but also reduced efficiency due to an increased gap between the two electrodes in the ESMG [19, 23, 24]. To avoid friction and wear, TENGs may use air as the dielectric but it is easy for air between the two frictionally charged surfaces to break down [25].

In this Letter we demonstrate that structural superlubricity [10, 11, 12, 13, 14] provides a revolutionary approach to producing ESMGs with superb performance. Figure 1a illustrates an elementary unit of our proposed superlubricity-based ESMGs, which consists of a SLIDER and a STATOR. The SLIDER is a thin electrode, Electrode 0, with a conductive smooth two-dimensional material as its bottom surface [26]. The

STATOR contains two dielectric films, Dielectrics 1 and 2, with a smooth surface on the top of Dielectric 1 and two parallel thin electrodes, Electrode 1 and Electrode 2, embedded in Dielectric 2. With superlubricity technology the contact between the SLIDER and the STATOR can form a superlubric joint [9, 10]. For simplicity in analysis, the three electrodes are set to be parallel at separating distances $d_1$ and $d_2$, respectively, and have the same rectangular shape with length $L$ and width $W$.

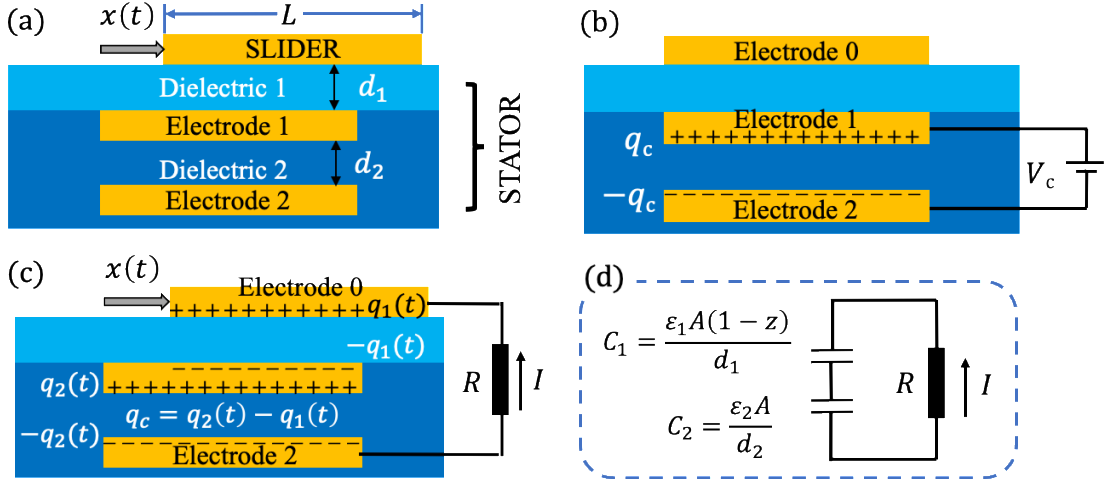

Figure 1. An elementary unit of the proposed superlubricity-based ESMGs: (a) The structure of the unit, consisting of a SLIDER and a STATOR contacted in a superlubric state. (b) The charging process. (c, d) The power generation process and its equivalent circuit diagram.

This ESMG model is prepared in two steps. First, we collect Electrodes 1 and 2 by a circuit (Figure 1b), apply a voltage $V_c$, and then turn off the circuit. This step leads to a storage of electric charges $\pm q_c$ in Electrode 1 and Electrode 2, with $q_c = C_2 V_c = \varepsilon_2 E_2 A$, where $C_2 = \varepsilon_2 A/d_2$ is the capacitance, $\varepsilon_2$ is the permittivity of Dielectric 2, $A = LW$ is the facing area of the two electrodes, and $E_2 = V_c/d_2$ is the electric field strength between the two electrodes. The main aim of this step is to store a charge in Electrode 1. To achieve this goal, one may also use Electrodes 0 and 1 temporarily as a capacitor for charging. Second, we collect Electrodes 0 and 2 by a circuit and an electric resistance $R$, as illustrated in Figure 1c. The equivalent serial capacitor and the circle are shown in Figure 1d.

A fundamental working model of the above ESMG is to slide Electrode 0 at a constant speed $v$ with respect to the STATOR. The changing serial capacity (Figure 1d) with time will therefore drive a time-dependent transferred charge $q(t)$ over the resistance $R$ to form a varying current $i(t)$, and thus works as an electrostatic generator. As detailed in SI, with this ESMG model we can obtain the implicit solutions of $q(t)$ and $i(t)$ through the following expressions for the dimensionless transferred charge $\Delta y(z) = q(t)/q_c$ and dimensionless current $y'(z) = \frac{dy(z)}{dz} = i(t)/q_c$:

$$y(z) = -\left[\frac{1}{1+\beta} + \alpha \int_0^z \frac{e^{\alpha\zeta}}{(1-\zeta)^{\alpha\beta}} d\zeta\right](1-z)^{\alpha\beta} e^{-\alpha z},$$
$$\Delta y(z) = y(z) - y(0) = y(z) + \frac{1}{1+\beta}, \qquad (1)$$

where $\varepsilon_1$ denotes the permittivity of Dielectric 1, $z = \frac{vt}{L} = \frac{t}{T}$ denotes the relative displacement or relative time, $T = \frac{L}{v}$ is the across-the-electrode time, and

$$\alpha = \frac{T}{C_2 R} = \frac{d_2 T}{\varepsilon_2 AR},$$
$$\beta = \frac{C_2}{C_1} = \frac{\varepsilon_2 d_1}{\varepsilon_1 d_2}, \qquad (2)$$

gives two key dimensionless parameters that characterize the performance of the studied model ESMG.

To validate the above elementary electrostatic generator based on two capacitors, we perform an experiment with an equivalent setup shown in Figure 2(a), in which the electrode (copper, lighter green) takes a half ring in the SLIDER and the STATOR, and the dielectric (acrylic oligomer, darker green) takes the other half. The rings for SLIDER and STATOR are coaxial, separated by a distance of ~0.25 mm, and in each experiment the SLIDER rotates at a constant angular speed maintained by a motor. An equivalent capacitor is connected with the STATOR to form a series of two capacitors. More details are given in SI. Figure 2(b) plots the measured currents versus time for four different rotatory speeds. As shown in Figure 2(c), the average values of the measured current peaks agree well with the theoretical predictions.

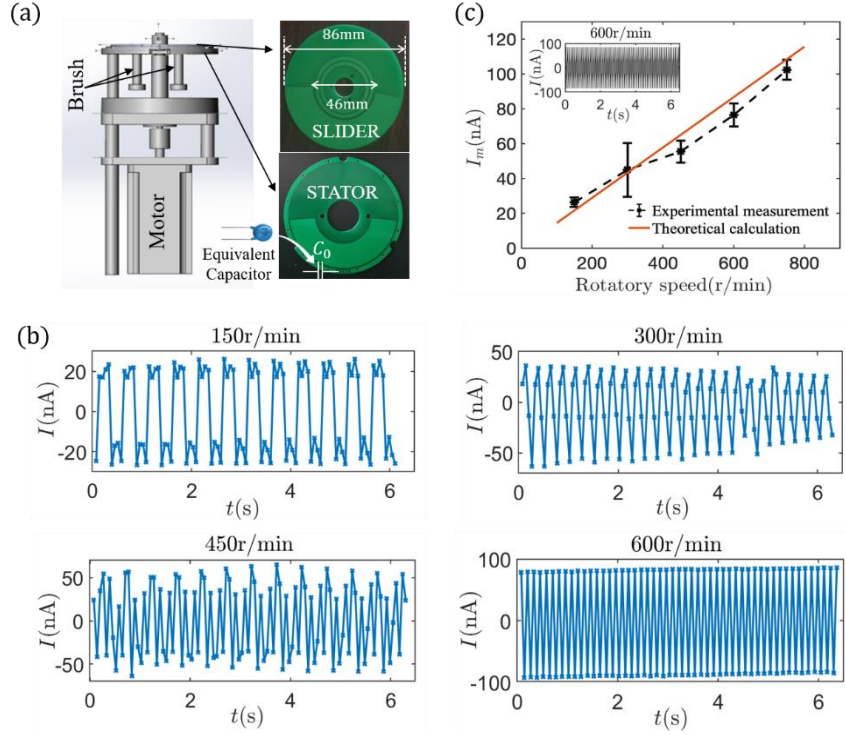

Figure 2. Experimental proof of concept of the proposed electrostatic generator, in which the sampling time interval is 0.06283±0.00046 s: (a) Experimental setup. (b) Measured instant currents $i(t)$ at four different angular speeds that correspond to four different sliding speeds. (c) Comparison with the theoretical predictions. The experimental error bars correspond to the current peaks at the first 20 rotating periods.

To optimize the design of our proposed ESMGs, we need more information of $q(t)$ and $i(t)$ as functions of time $t$ and their dependence upon the structural and material parameters. A detailed analysis on these functions is given in SI. From the numerical solution of $\Delta y$ in the phase space $(z, \alpha)$ as detailed in SI, we can see that for each given $\alpha$, $\Delta y$ is monotonically increasing with $z$ and reaches its maximum, $\Delta y_{\max}$, at $z = 1$. Further, $\Delta y_{\max} = \Delta y|_{z=1}$ as a function of $\alpha$ becomes a plateau (constant) for all sufficiently large $\alpha$. Similar results hold for $y'$. In fact, we can prove (see SI) that the plateau values with respect to any given finite $\beta$ are

$$\Delta y_{\mathrm{pt}} = \lim_{\alpha \to \infty} \Delta y|_{z=1} = \frac{1}{1+\beta},$$
$$y'_{\mathrm{pt}} = \lim_{\alpha \to \infty} \Delta y|_{z=1} = \frac{1}{\beta}. \tag{3}$$

Practically, since permittivities of dielectrics are very small (typically several times $\varepsilon_0 = 8.8542 \times 10^{-12}$ in a vacuum), in most cases $\alpha$ is much larger than 1 so that (3) holds. For example, if using SiO$_2$ as Dielectric 2 ($\varepsilon_2 = 4.9\varepsilon_0$) and taking $\frac{W}{d_2} = 1000$, $v = 1m/s$, and $R = 1000\Omega$, then we have $\alpha = 2.3 \times 10^4$.

Since both the transferred charge, $q = q_c\Delta y$, and the current, $i = q_c y'$, are proportional to the capacitor charge $q_c$, to achieve the maximal possible $q$ and $i$, we can first explore the maximal allowed $q_c$ before breaking down any of the two dielectric films. According to $q_c = \varepsilon_2 E_2 A$, the maximal allowed $q_c$, $q_{c,max}$, must satisfy $q_{c,max} \leq \sigma_{2cr}A$, where $E_{2cr}$ denotes the breakdown electric strength of Dielectric 2, and $\sigma_{2cr} = \varepsilon_2 E_{2cr}$ corresponds to the maximal allowed charge density. Again with silicon oxide as Dielectric 2 ($E_{2cr} \approx 10^9 V/m$) for example, to achieve $q_{c,max}$ we need only to apply a favorably low voltage $V_c = 10V$ if the thickness of the dielectric between Electrodes 0 and 2 can be set at $d_2 = 10nm$ (Note: the thinnest films reported through coating are about 0.5 nm [27]); in comparison, if $d_2 = 100$ μm, which is a typical case in conventional ESMGs, then one has to apply an extremely high voltage $V_c = 100,000\ V$ in order to achieve the maximum $q_{c,max} = \sigma_{2cr}A$. This means that $q_{c,max}$ is impractical to achieve for most convenient ESMGs.

We have to avoid electric breakdown of the dielectric media in not only the capacitor charging process (Step 2) but also the electricity generating process. Therefore, we need to study the electric field change during the generating process that corresponds to the sliding process for the entire period $0 \leq t \leq T$. In this process, it is easy to find that the electric field strength $E_2$ in Dielectric 2 automatically satisfies $E_2 < E_{2cr}$ as long as Dielectric 2 is not broken down in the charging process (see SI). However, the electric field $E_1$ in Dielectric 1 may reach the breakdown limit of Dielectric 1. To avoid this, the maximal allowed capacity charge has to satisfy $q_{c,max} \leq k\varepsilon_1 E_{1cr}A$, where $k$ is a function of $\alpha$ and $\beta$, and $E_{1cr}$ is the breakdown electric strength of Dielectric 1. As detailed in SI, again we find that for sufficiently large $\alpha$ ($> 10$), $k$ is virtually equal to β. Therefore, the maximal allowed capacity charge $q_{c,max}$ is equal to $k_m\varepsilon_2 E_{2cr}A$, where $k_m = \min(1, \frac{\beta\varepsilon_1 E_{1cr}}{\varepsilon_2 E_{2cr}})$ for sufficiently large α. We further find (see SI) that the optimized β is $\beta_{opt} = \frac{\varepsilon_2 E_{2cr}}{\varepsilon_1 E_{1cr}}$, corresponding to

$$V_c = E_{1cr}d_1 = E_{2cr}d_2. \tag{4}$$

The maximal allowed transferred charge $q_{max}$, maximal allowed instant current $i_{max}$, and maximal allowed average power $p_{max}$ can be achieved as follows:

$$\begin{aligned} q_{max} &= \frac{1}{1+\beta_{opt}} q_{c,max}, \\ i_{max} &= \frac{1}{\beta_{opt}T} q_{c,max}, \\ p_{max} &= \frac{1}{3}R \frac{3+3\beta_{opt}+\beta_{opt}^2}{(1+\beta_{opt})^3}\left(\frac{q_{c,max}}{T}\right)^2, \end{aligned} \tag{5}$$

where $q_{c,max} = \varepsilon_2 E_{2cr} A$.

Superlubricity-Based ESMGs offer quite a few significant advantages over the conventional ones. First, through shrinking the thicknesses $d_1$ and $d_2$ into nanoscales, we can achieve the theoretically maximal permissive capacity charge $q_{c,max} = \varepsilon_2 E_{2cr} A$ and the maximal possible values for all three major performance indices ($q_{max}$, $i_{max}$, and $p_{max}$) by using a very low voltage $V_c = E_{2cr}d_2$. This is critical in practice, in particular for healthcare applications where a high voltage is typically undesirable and needs to be avoided.

Second, according to (5), all three performance indices decrease with $\beta_{opt}$ in the form $\sim \beta_{opt}^{-1}$ for large $\beta_{opt}$. In conventional ESMGs, a gas or even a vacuum layer has to be adopted as Dielectric 1 in order to avoid friction and wear. As a result, $\beta_{opt}$ is in the order of $10^2$ or even larger [19, 23, 24]. In addition, as just discussed, to achieve $q_{c,max}$ in conventional ESMGs requires the application of very high voltages $V_c$. In other words, the acceptable $q_c$ would be orders lower than $q_{c,max}$. In sum, the performance indices of superlubricity-based ESMGs can be at least three orders of magnitude higher than those of conventional ESMGs.

Third, for sufficiently large $\alpha$, the maximal external force required for the best possible performance shown in (5) can be solved as:

$$F_{e,max} = \frac{1}{2}\varepsilon_1 E_{1cr} E_{2cr} W d_2. \tag{6}$$

Therefore, the required external force can become very low if we use the dielectric film with superlubricity-endowed nanoscale thickness $d_2$. For instance, when $L = 10$ μm and both Dielectrics 1 and 2 are silicon oxide ($E_{1cr} = E_{2cr} \approx 10^9$ V/m and $\varepsilon_1 = \varepsilon_2 =$

$4.9\varepsilon_0$), Figure 3a displays the maximum external force needed to slide with respect to different widths $W$ (with $d_1$ satisfying $\beta = \beta_{\text{opt}} = 1$) when the resistance is $R = 10^5 \Omega$. From this plot we observe an important feature: the external force needed for driving the SLIDER can be reduced to the magnitude of μN when $d_2$ is reduced to the nanoscale. At the same time, because of superlubricity, the mechanical-to-electrical efficiency can be very high (>97%), as shown in Figure 3b.

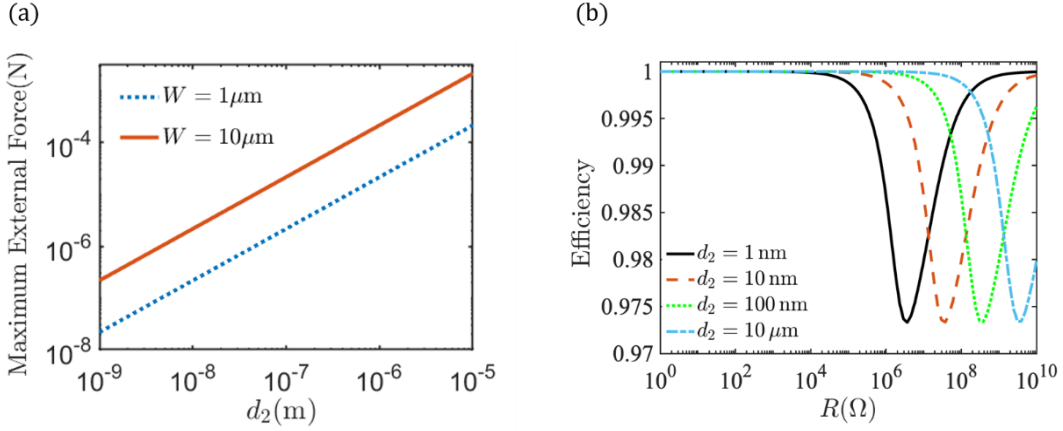

Figure 3. The maximum external force and efficiency: (a) Dependence of maximum external force required to achieve the best possible performance upon the dielectric thickness $d_2$ for different widths $W$. (b) Dependence of efficiency upon the resistance $R$ for different dielectric thicknesses $d_2$ (with $d_1$ satisfying $\beta = \beta_{\text{opt}} = 1$).

Fourth, the structure shown in Figure 1 can be easily and massively fabricated by using micro-machining processes. This feature is particularly important for commercialization of ESMGs.

Fifth, it is revealed by (6) that reducing the across-the-electrode time $T = L/v$, through shrinking $L$ or increasing $v$, can allow higher $i_{\max}$ and $p_{\max}$. The former approach is practically important since the current size limit (~10 $\mu m$) of realized structural superlubricity becomes an advantage in achieving high $i_{\max}$ and $p_{\max}$ for superlubricity-based ESMGs.

Sixth, the electric breakdown strengths, $E_{\text{cr}}$, of a number of solid dielectric films exhibit an increasing tendency with shrinking film thickness as the film falls in the nanoscale [28, 29]. Therefore, with smaller thickness we can obtain improved breakdown strength of the solid dielectric and thereby further increase the maximum allowed charge, $q_{c,\max}$.

Before concluding this Letter, we show two simple, and thus relatively easy, setups of superlubricity-based ESMGs. First, we suggest a rotatory ESMG as illustrated in Figures 4a and 4b. In the ESMG, a number ($n$) of equiseparated, same-sized, and circular sector-shaped thin electrodes (the top orange flakes) connected by a ring conductor (ignored for simplicity) form the SLIDER; and two layers containing the same number of the same electrodes, connected by an inner conductive wire, are used instead of Electrode 1 and Electrode 2 in Figure 1 (and E1 and E2 in Figure 4b). For this rotatory ESMG, if we suppose that Dielectrics 1 and 2 are the same (D1 and D2 in Figure 4b), then from (4) by setting $\beta_{\text{opt}} = 1$ we can obtain the maximal allowed instant transferred charge, instant current, and average power, in the following forms:

$$Q_{\max} = \frac{1}{2} Q_{c,\max},$$
$$I_{\max} = T^{-1} Q_{c,\max}, \tag{7}$$
$$P_{\max} = \frac{7}{24} R\left(T^{-1} Q_{c,\max}\right)^2,$$

where $R$ denotes the total resistance, and $Q_{c,\max} = n q_{c,\max}$ is the maximal allowed total capacity charge.

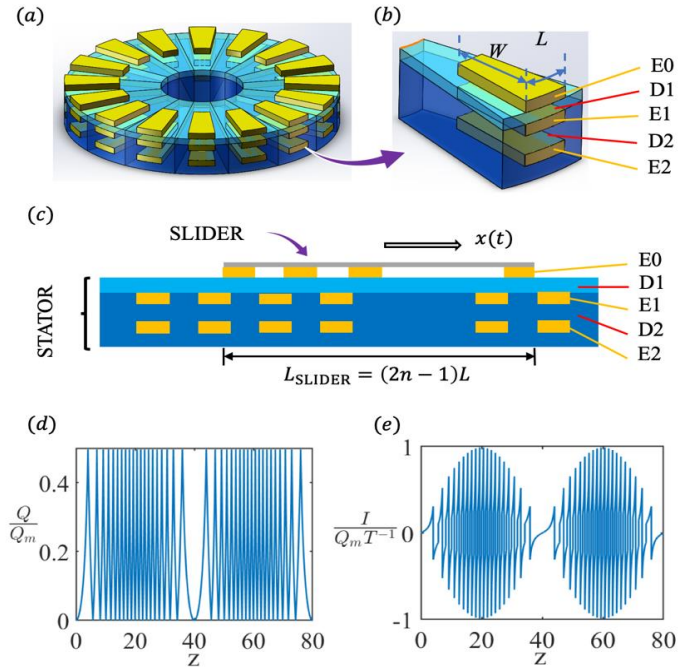

Figure 4. Two simple setups of superlubricity-based EMSG: ($a$) The schematic diagram of the rotatory form superlubricity-based ESMG. ($b$) Partial, enlarged view of the array unit, where E's denote

Electrodes, and D's indicate dielectrics. (*c*) The schematic diagram of the linear array superlubricity-based ESMG. (*d, e*) The corresponding time-dependent dimensionless transferred charge and instant current of the linear array.

The second setup of the ESMG is illustrated in Figure 4c, where the SLIDER consists of a linear array of equi-spaced, same-sized, and square-shaped electrodes connected by a conductive wire (ignored for simplicity). The total length of the SLIDER is $L_{\text{SLIDER}} = (2n-1)L$, where *n* denotes the number of units in the array. This is much like an elementary model for vibrational energy harvesters [1, 23, 24, 30], where the sliding displacement $x(t)$ can be random. For a quantitative understanding, we analyze a simple overall oscillatory form: $x(t) = \frac{L_\text{m}}{2}(1 - \cos\frac{2v_{\max}t}{L_\text{m}})$, where $L_\text{m}$ denotes the maximum displacement distance of the SLIDER and $v_{\max}$ the maximum instant speed. The maximal instant transferred charge and instant current have exactly the same forms as in $(6)_{1,2}$ (for $\beta_{opt} = 1$) while the average power is slightly lower than that given in $(6)_3$. For example, as $L_m/L = 40$, the result of $P_{\max}$ is the coefficient 0.4965 times the right-hand side of $(6)_3$. The corresponding time-dependent dimensionless transferred charge $\Delta y$ and instant current $dy/dz$ are shown in Figures 4d and 4e for two oscillatory periods.

In conclusion, we propose an elementary superlubricity-based ESMG and present explicit results of three major performance indices of ESMGs: the maximal possible instant transferred charge $q_{\max}$, maximal possible instant current $i_{\max}$, and maximal possible average power $p_{\max}$. The results show that superlubricity-based ESMGs enjoy great advantages (at least three orders of magnitude higher) over conventional ones on all three performance indices. Meanwhile, the voltages and external forces needed can be at least three orders of magnitude lower. Therefore, commercialization of ESMGs can be expected to a reality soon because of superlubricity, which will then lead to their broad applications.